 \documentclass[preprint,showpacs,preprintnumbers,amsmath,amssymb]{revtex4}

\usepackage{epsf}
\usepackage{graphicx}
\usepackage{dcolumn}
\usepackage{bm}
\def\be{\begin{equation}}
\def\ee{\end{equation}}
\def\bea{\begin{eqnarray}}
\def\eea{\end{eqnarray}}

\def\nonu{\nonumber}

\newlength{\dinwidth}
\newlength{\dinmargin}
\begin{document}
 \tighten
\vskip 3cm

\

\preprint{SU-ITP-01/40, ~SLAC-PUB-9016, ~hep-th/0110089}
\title{\Large\bf Gauged Supergravities, 
de Sitter Space and Cosmology  \\ \phantom{} }

\author{\bf Renata Kallosh,$^1$ Andrei Linde,$^1$
 Sergey Prokushkin$^1$  and Marina Shmakova$^2$ }

\affiliation{ {$^1$Department
  of Physics, Stanford University, Stanford, CA 94305-4060,
USA}    }   

\affiliation{$^2$CIPA, 366 Cambridge Avenue,
 Palo Alto, CA 94306, and  SLAC,
 Stanford University, Stanford, CA 94309  }

\date{October 9, 2001 \\ \phantom{}}
 
 {
\begin{abstract}We have studied  scalar potentials 
$V$ of gauged   N=8,4,2 supergravities in d=4. Extrema
 of these potentials may correspond to de Sitter, 
anti de Sitter and Minkowski vacua.
 All de Sitter extrema that we have studied correspond to 
unstable maximum/saddle points with negative curvature 
$|V''|=2V$ for the fields canonically normalized at the
 extremum. This  is equivalent to the relation 
$|m^2| =|R|/2 = 6H^2$ for the tachyonic 
mass $m$, the curvature scalar $R$, and the Hubble constant H.  
 This  prevents the  use of de Sitter extrema for slow-roll 
inflation in the early universe, which would require  
$|V''| \ll V $. Moreover, in all models that we were able to 
analyse the potential is unbounded from below.  On the other hand, 
barring the question how realistic such 
models could be, one can use them for the description 
of the accelerated expansion of the universe at the  present epoch. 
This is related to a novel 
possibility of the fast-roll inflation with $|V''|\gtrsim V $. 
We also display some  potentials  that have flat directions with 
vanishing cosmological constant, and discuss their possible 
 cosmological implications.
\end{abstract}}

\maketitle


\section{Introduction}

Recent cosmological observations based on the study of supernova
 \cite{supernova} and of the anisotropy of the cosmic microwave
 background radiation \cite{CMB} suggest that soon after the big
 bang our universe  experienced a stage of a very rapid 
 accelerated expansion (inflation) \cite{Guth,New,chaot,hybrid}. 
Moreover, observations indicate that few billion years after 
the big bang  the universe entered a second stage of accelerated
 expansion. The rate of acceleration  now is many orders of 
magnitude smaller than during the stage of inflation in the 
early universe. For a discussion of the recent cosmological 
observations and their theoretical interpretation 
see e.g. \cite{astroreviews}.

Most of the inflationary models are based on the assumption  
that the energy-momentum tensor during inflation is dominated
 by potential energy density of a scalar field,  
$T_{\mu\nu} \approx g_{\mu\nu} V(\phi)$ with $V >0$  
\cite{book,LythLiddle}. This means, in particular, 
that $\dot\phi^2/2 \ll V(\phi)$. The limiting case 
$\dot\phi =0$ corresponds to de Sitter space with a 
positive cosmological constant. The  current cosmological 
acceleration can be explained either by a positive vacuum
 energy $V$ (cosmological constant)  or by a slowly rolling 
scalar field in a near de Sitter background with 
$\dot\phi^2/2 \ll V(\phi)$ (quintessence) \cite{quint}. 

Much of the recent progress in theory of all fundamental 
interactions is related to supersymmetry. Therefore it 
would be very desirable to find  de Sitter  or 
near-de Sitter solutions in supersymmetric theories. 
In N=1 theories  many choices are available and therefore 
the motivation for a particular choice has to come from 
a more fundamental theory, either directly from M/string 
theory in 11/10 dimensions or  perhaps from an effective 
4d theory with higher supersymmetry.

This  was recently discussed in the context of 4d N=2 
supersymmetry in \cite{Kallosh:2001tm} where a hybrid 
hypersymmetric  model of inflation/acceleration was 
proposed. On the other hand, it may be difficult 
to obtain 4d de Sitter space directly from M/string theory. 
The problems have been recently reviewed in 
\cite{Hull:2001ii,Spradlin:2001pw}.

The basic problem seems to originate from the  
compactification of M/string theory on internal space
 with the finite volume. However, some  
4d N=8,4,2 gauged supergravity theories  are known to have 
 de Sitter solutions with spontaneous breaking of supersymmetry 
\cite{Gates:1983ct}-\cite{deWit:1984pk}.
 These versions of 4d supergravity are related to 11d 
supergravity with a non-compact internal 7d space 
\cite{Hull:1988jw}.  M-theory has some solutions with warped
 products of de Sitter space with hyperbolic spaces and 
 generalized cylinders,  $dS_4\times {\cal H}^{p,q,r}$.
 For conceptual problems associated with non-compact
 internal space and attempts to overcome them we refer
 to \cite{Hull:2001ii} and references therein. 

Here we will take a simple attitude to potentials of gauged 
supergravities: we will consider  only 4d theories where all
 kinetic terms have correct sign and therefore these theories
 from the 4d perspective do not have any conceptual problems
 whatsoever.  We will study the properties  of their potentials 
with regard to cosmology. If we find any potentials interesting
 for cosmology, we may come back to the problems of accommodating 
these 4d supergravity models in the framework of  M-theory 
with non-compact internal space.

The physical motivation for our study of potentials of 
supergravity originates in cosmology. A standard 
{\it slow-roll inflation} resulting in many e-foldings 
and scale independence of the spectrum will be reviewed. 
A new concept of {\it fast-roll inflation} will be proposed.
 It is partially motivated by the properties of supergravity 
potentials and also by the present epoch acceleration 
with small number of e-foldings.

We start with reminding some well known facts about the
 relation between de Sitter space and expanding universe 
in Section \ref{dS}. Section \ref{fast} explains a standard 
{\it slow-roll inflation} and introduces a new concept of
 a {\it fast-roll inflation}. This defines the properties 
of the potentials interesting for cosmology.

In Sections \ref{N=8} and \ref{N=4} we study the potentials
 of N=8 and N=4,  2 gauged supergravities. In Section 
\ref{flat} we discuss some potentials of N=8 supergravity
 which have interesting Minkowski vacua with flat 
directions. We discuss the main results and perspectives
 in Conclusions.


\section{\label{dS}de Sitter versus anti de Sitter}

Anti de Sitter space was at full attention of high-energy
 physics community for the last 10 years from the time when 
it was realized that  near horizon the geometry of stringy
 BPS black holes and branes tends to anti de Sitter space. 
It was known also for a very long time that the potentials 
of gauged supergravities have anti de Sitter critical points 
which are at the top of the inverted potentials and correspond 
to the maximum of these potentials and the  tachyons are
 present. Still when the negative (mass)$^2$ is limited
 in a certain way corresponding to Breitenlohner-Freedman
 bound, the anti de Sitter solution is stable. 
For $adS_4$ the bound is 
\begin{equation}
 m^2\geq -{9\over 4 } H^2,
\end{equation} 
where $H^{-1}$ defines the radius of the throat of the 
  hyperboloid in the flat 5d space
\begin{equation}
 X^2_0 -X^2_1-X^2_2 -X^2_3+ X^2_4 = H^{-2} \ .
\end{equation} 
The hyperboloid in the 5d flat space defining  de Sitter space  $dS_4$ is
\bea
X^2_0 -X^2_1-X^2_2 -X^2_3- X^2_4 = -H^{-2} \ .
\label{deSitter}\eea 
The difference between the two hyperboloids is that the right
 hand side has an opposite sign. Besides, the $adS$ hyperboloid
 has two `time directions' whereas $dS$ one has only one
 `time direction'. The symmetry group for the $adS$ hyperboloid 
is $SO(3,2)$ whereas for the $dS$ hyperboloid it is $SO(4,1)$.

To explain shortly the relation of de Sitter space to cosmology 
we will first present here the Friedmann-Robertson-Walker (FRW) 
metric describing the expanding Friedmann universe:
\bea
ds^2= dt^2- a^2(t)\left[{dr^2\over 1-kr^2} 
+r^2(d\theta^2 + \sin^2\theta d\phi^2)  \right],
\eea
where $k=\pm 1, 0$ for a closed, open or flat expanding 
Friedmann universe, respectively. Here $a(t)$ is the 
time-dependent scale factor. For a spatially flat
 universe with $k=0$ the metric can be represented in a form
\bea
ds^2= dt^2- a^2(t)d\vec x^2 \  .
\eea
Closed Friedmann universe with $k=1$ can be 
represented in the form
\bea
ds^2= dt^2- a^2(t)\left[d\chi^2 + \sin^2( \chi) 
(d\theta^2 + \sin^2\theta d\phi^2)  \right]\ ,
 \qquad 0\leq \chi\leq 2\pi \ .
\eea
Finally, an open Friedmann universe with $k=-1$
 can be represented in the form
\bea
ds^2= dt^2- a^2(t)\left[d\chi^2 + \sinh^2( \chi) 
(d\theta^2 + \sin^2\theta d\phi^2)  \right]\ ,
 \qquad 0\leq \chi\leq \infty \ .
\eea
The relation between the metric of expanding universe 
and  de Sitter space is the following. First, one can 
consider the hyperboloid (\ref{deSitter}) in a coordinate 
system which spans the half of the hyperboloid with 
$X_0 + X_4>0$. The choice 
$$X_0= H^{-1} \sinh Ht + {1\over 2}H e^{Ht} \vec x^2 ,
 \qquad X_4= H^{-1} \cosh Ht - {1\over 2}H e^{Ht},
 \qquad X_i = e^{Ht}x_i, i=1,2,3$$ 
leads to the metric 
\bea
ds^2= dt^2- e^{2Ht}d\vec x^2   \ .
\eea
This is a FRW spatially flat metric with the exponential
 scale factor $a(t)= e^{Ht}$. 
We may choose the  coordinates which span the entire
 hyperboloid : 
$$X_0= H^{-1} \sinh Ht\ , \qquad X_1= H^{-1}
 \cosh Ht \cos \chi\ , \qquad X_2= H^{-1} \cosh Ht\,
 \sin \chi\, \cos \theta\ , $$
$$ X_3= H^{-1} \cosh Ht\, \sin \chi\,
 \sin \theta\, \cos \phi\ , \qquad X_4= H^{-1}
 \cosh Ht \sin \chi\, \sin \theta\, 
\sin \phi \ ,$$
which results in the metric
\bea
ds^2= dt^2- H^{-2} \cosh^2 Ht\left[d\chi^2 + \sin^2( \chi)
 (d\theta^2 + \sin^2\theta d\phi^2)  \right]\ ,
 \qquad 0\leq \chi\leq 2\pi \ .
\eea
This is a FRW metric of the closed Friedmann universe
 with the exponential scale factor $a(t)= H^{-1}\cosh Ht$.
 An open Friedmann universe appears from de Sitter space if 
the coordinate system is chosen as follows:
$$X_0= H^{-1} \sinh Ht\, \cosh \chi, \qquad X_1= H^{-1}
 \cosh Ht , \qquad X_2= H^{-1} \sinh Ht\, 
\sinh  \chi\, \cos \theta\ , $$
$$ X_3= H^{-1} \sinh Ht\, \sinh \chi\,
 \sin \theta\, \cos \phi , 
\qquad X_4= H^{-1} \sinh Ht\, 
\sinh \chi\, 
\sin \theta\, \sin \phi \ .$$
These coordinates do not cover the entire hyperboloid.
 The  metric is
\bea
ds^2= dt^2- H^{-2} \sinh^2 Ht\left[d\chi^2
 + \sinh^2( \chi) (d\theta^2 +
 \sin^2\theta d\phi^2)  \right]\ ,
 \qquad 0\leq \chi\leq \infty \ .
\eea
This is a FRW metric of the open Friedmann universe 
with the exponential scale factor $a(t)= H^{-1}\sinh Ht$.


\section{\label{fast} de Sitter space, slow-roll inflation 
and fast-roll inflation}
De Sitter space has a direct relation to inflationary 
cosmology \cite{Guth,New,chaot,hybrid,book,LythLiddle}, 
which we will briefly describe now. 
Consider  a theory of a scalar field $\phi$ with 
potential $V(\phi)$. The Friedmann equation for the scale 
factor of the universe looks as follows:
\bea\label{freed}
\left({\dot a\over a}\right)^2 + {k\over a^2} = {\rho(\phi)\over 3}  
\eea
in units $M_p = 1$. Here $k = \pm 1,0$ for a closed,
 open or flat universe correspondingly,  
$\rho(\phi) = V(\phi) + \dot\phi^2/2 
+ (\partial_i \phi)^2/2$ is the energy density 
of the scalar field.

Let us assume first that 
$V(\phi) = V = const > 0$, and the field 
$\phi$ is constant and homogeneous, 
 $\dot\phi =  \partial_i \phi = 0$.
Then   
\bea
\left({\dot a\over a}\right)^2 
+ {k\over a^2} = {V\over 3}. 
\eea
The solutions of this  equation describe 
de Sitter space with
\bea
H  = \sqrt{V\over 3}. 
\eea
Note that at very large times $t \rightarrow \infty$ 
all 3 types of de Sitter metric for the flat, closed 
and open universe   lead to the same exponential
scale factor
\bea
e^{Ht} \ , \qquad \cosh Ht \rightarrow e^{Ht}\  ,
 \qquad \sinh Ht \rightarrow e^{Ht} \ .
\eea
Therefore  we will concentrate on the simplest 
case of the flat universe with $a(t) = a(0)\, e^{Ht}$.
 Flatness of the universe is a standard prediction 
 of most of the inflationary models, where typically
 the term ${k\over a^2}$ can be neglected as compared 
with $\rho/3$ after inflation \cite{Guth,New,chaot,
hybrid,book,LythLiddle}. This means that  inflationary 
theory predicts that our universe at present cannot be 
in  the anti-de Sitter regime, because Eq. (\ref{freed})
 does not have any solutions for  $\rho <0$ in the
 flat universe.

De Sitter space can describe late stages of the evolution 
of our universe if the universe has nonvanishing vacuum 
energy $V \lesssim 10^{-120} M_p^4$. However, in order
 to use de Sitter-like stages for a description of 
inflation in the early universe  one should find how 
de Sitter  stage ends and the usual hot Friedmann
 universe emerges.

One possibility is to study models where the potential
 $V(\phi)$ has a very flat maximum, as in the new 
inflation scenario \cite{New}. Consider, for example,
 the model where $V(\phi)$ has a maximum at $\phi = 0$, 
such that in the vicinity of the maximum 
\bea
V(\phi) = V_0 - {m^2\phi^2\over 2} \ .
\eea
Suppose that initially the field $\phi = \phi_0$ was
 homogeneous, small, and had small velocity, so that 
${m^2\phi^2/2}, \dot\phi^2/2 \ll V_0$. Then the Hubble
 constant $H^2 = \rho/3 = {1\over 3}(V_0 
- {m^2\phi^2/2 + \dot\phi^2/2})$ practically 
did not change until the field rolled down to 
the point with $ m^2\phi_1^2 \sim V_0$. 
Therefore the universe continues expanding 
as $e^{Ht}$ until the field $\phi$ rolls from $\phi_0$ to 
\bea
 \phi_1 \sim \sqrt{V_0}/m  \ .
\eea 

The motion of a homogeneous field $\phi$ is described 
by equation
\bea\label{genequation}
\ddot\phi + 3H\dot\phi = - V' = m^2\phi \ .
\eea
Now let us assume that $|V''| = |m^2| \ll H^2 = V/3$. 
In this case one can show that the field moves very 
slowly, so that one can neglect $\ddot\phi$ as compared 
to $3H\dot\phi$, and the growing solution for the field 
$\phi$ is given by
$
\phi =\phi_0 \exp{m^2 t\over 3H} \ .
$
This is the standard slow-roll solution for the scalar 
field during inflation. The slow-roll regime continues
 at least until the field rolls down below $\phi_1$, 
i.e. during the time 
\bea
\Delta t = {3H \over m^2} \log {\phi_1\over\phi_0} .
\eea
This leads to inflation by a factor of
\bea
e^{H\Delta t} \sim 
\left({\phi_1\over\phi_0}\right)^{3H^2/m^2} 
\sim \left({\phi_1\over\phi_0}\right)^{V/|V''|}  \ .
\eea

Thus one can obtain an exponentially large 
degree  of inflation for 
\bea
 |V''|\ll |V|  ,
\eea
which is one of the two well-known inflationary 
slow-roll conditions: $\eta = {|V''|\over |V|} \ll 1$ and 
  $\epsilon = {1\over 2} \left({V'\over V}\right)^2 \ll 1$ 
\cite{LythLiddle}. The last condition is automatically 
satisfied at the top of the effective potential.

The slow-roll conditions serve two purposes: They make 
the total expansion of the universe during the stage of 
inflation very large, $e^{H\Delta t} \sim 
\left({\phi_1\over\phi_0}\right)^{|V|/|V''|}$, 
and they ensure that the spectrum of adiabatic density 
perturbations produced during inflation  is almost 
scale-independent. These density perturbations are produced
 due to quantum effects during inflation \cite{pert};
 they are playing a critical role in the subsequent 
process of formation of the large-scale structure 
of the universe \cite{book,LythLiddle}. The deviation 
from scale-independence (flatness) of the spectrum is
 characterized by the factor $|n-1| = |2{|V''|\over |V|} 
-3\left({V'\over V}\right)^2|$ \cite{LythLiddle}. Recent
 observations of anisotropy of the cosmic microwave 
background radiation suggest that $|n-1| \lesssim 0.1$ 
\cite{CMB}, which implies that  
\bea
|V''|\lesssim 0.05  V \ .
\eea
if the perturbations that we observe were 
produced during inflation when the field
 $\phi$ was near the top of the potential.

The slow-roll condition $ |V''|\ll V$  implies 
that $|m^2| \ll V_0$, and, consequently,
\bea
\phi_1 \gg 1 \  
\eea
in Planck units (i.e. $\phi_1 \gg M_p$). This is very 
similar to the  standard requirement that appears in the
 simplest models of chaotic inflation with
 $V(\phi) \sim \phi^n$ \cite{chaot,book}. 
To avoid this requirement and still have 
slow-roll inflation one would need to make
 the potential very flat at the top, 
and very curved near the minimum of the potential,
 as in the original version of the new
 inflation scenario \cite{New}, or as in 
the hybrid inflation scenario \cite{hybrid}.

One of the results of our paper is that the slow-roll 
condition $V''\ll  V$ is not satisfied near any 
of the extrema of the potentials with $V >0$ in 
N=8 gauged supergravity that have been studied in 
the literature. As we will show, in all of the 
models of N=8 gauged supergravity studied so
 far one has $V''= 2  V$. Thus, in all known 
cases these models do not support slow-roll i
nflation near the extrema of the corresponding 
potentials.

Is it possible to have a {\it fast-roll inflation} 
with $|V''| = |m^2| \geq H^2 = V/3$? In our 
investigation of this question we have found,
 much to our own surprise, that the answer to
 this question is positive.

For simplicity we will consider first  the 
limiting case $|m^2| \gg H^2$. In this case
 one can neglect the term $3H\dot\phi$ in 
the equation for the field $\phi$.  
Then the growing solution becomes
\bea
\phi =\phi_0 e^{m t} \ .
\eea
The duration of rolling from 
$\phi_0$ to $\phi_1$ is given by
\bea
\Delta t = m^{-1} \log {\phi_1\over\phi_0} .
\eea
Until the field rolls down to $\phi_1$ the 
energy density remains dominated by $V_0$.
 This leads to  inflation by a factor of
\bea \label{expansion}
e^{H\Delta t} \sim \left({\phi_1\over\phi_0}\right)^{H/m}  \ .
\eea
Usually one does not expect the ratio 
${\phi_1\over\phi_0}$ to be exponentially large, 
and therefore one could think that for $H/m <1$ the 
duration of inflation of this kind must be rather
 insignificant. Also, no long-wavelength perturbations 
of metric are generated  in the regime $m > H$ 
by the standard inflationary mechanism. That is 
why  the possibility of a   fast-roll inflation  
with $m \geq H$ has not been thoroughly studied 
in the literature. 

Meanwhile fast-roll inflation can be quite interesting, 
at least for a marginally fast-rolling regime with 
$m \sim H$ (i.e. $\sqrt{|V''|} \sim \sqrt V$). 
This is  the regime that we will often encounter 
in our investigation of extrema of the potentials 
in N=8 gauged supergravity.

First of all, let us note that the initial value of the
 field $\phi$ can be quite small. Formally, one may 
have $\phi_0 = 0$. In this case  the factor 
${\phi_1\over\phi_0}$ in Eq. (\ref{expansion})
 can be indefinitely large. In reality, $\phi_0$ 
in this equation cannot be taken much smaller than 
the level of quantum fluctuations with momenta $k<m$, 
since such fluctuations also experience exponential
 growth, even in the absence of a homogeneous field $\phi_0$: 
$\delta\phi_k(t) \sim \delta\phi_k(0)e^{\sqrt{m^2-k^2}\, t}$.
A typical initial amplitude of all quantum fluctuations 
with $k < m$ participating in the exponential growth 
of the field $\phi$ can be estimated by 
$\delta\phi \sim C m$, where $C = O(10)$ \cite{FKL}.
 A typical time it takes for this field 
to grow up to $\phi_1$ is given by \cite{FKL}
\bea \label{constC}
\Delta t \sim m^{-1} \log {C \phi_1\over m} .
\eea
This leads to  inflation by a factor of
\bea  \label{smallm1} 
e^{H\Delta t} \sim \left({10\phi_1\over m}\right)^{H/m}  \ .
\eea

Now that we have studied two limiting case, let us study
 a more general regime where $m$ and $H$ can be of 
the same order. To study this problem one should 
look for solutions of Eq. (\ref{genequation}) 
in the form $\phi = \phi_0 e^{i\omega t}$. 
This yields
\bea    
\omega =i \left({3H\over 2}
 \pm \sqrt{{9H^2\over 4} +m^2} \right). 
\eea
The solution with the sign - corresponds to the exponentially 
growing  solution  
\bea    
\phi =\phi_0 \exp \left [\left(Ht\cdot F(m^2/H^2) \right)\right] , 
\eea
where 
\bea    
F({m^2/H^2}) = \sqrt{{9\over 4} + {m^2\over H^2}} - {3\over 2} \ . 
\eea
This immediately gives us the general result for the total 
expansion of the universe during inflation near the 
maximum of the  potential:
\bea    
e^{H\Delta t} \sim \left(
{\phi_1\over \phi_0}\right)^{1/F}.
\eea
One can easily check that this result coincides 
with our previously obtained results in the limiting 
cases $m\gg H$ and $m\ll H$.

As an example, consider first the potentials with $m = H$. 
In this case one has $F(1) = 0.3$. In the theories with 
$m \sim H$ one has  $\phi_1 \sim M_p = 1$, so for 
$\phi_0 \sim 10 m^{-1}$ (i.e. for the initial value of 
the field provided by quantum fluctuations \cite{FKL})  one has 
\bea \label{smallm}
e^{H\Delta t} \sim  \left({10M_p\over m}\right)^{3.3}  \ .
\eea
Clearly, this number can be quite significant.

To be more specific, consider the possibility that such 
models can be responsible for the present stage of accelerated
 expansion of the universe with the Hubble constant
 $H \sim 10^{-60}M_p$. Then inflation in an unstable 
state close to the maximum of the potential in such a 
theory can lead to expansion of the universe by a 
factor that can be as large as 
\bea
e^{H\Delta t}  \sim \left({10^{61}}\right)^{3.3} 
\sim 10^{200} \sim e^{460} \ .
\eea
This is more than sufficient to explain the observed 
single e-folding of accelerated expansion of the 
universe at the present epoch.

Meanwhile if one takes $m \sim 10^2$ GeV 
$\sim 10^{-16} M_p$, which corresponds to the electroweak 
scale, one can obtain fast-roll inflation by a factor of 
\bea
e^{H\Delta t} \sim  \left({10^{17}}\right)^{3.3} 
\sim 10^{56}\sim e^{130}  \ .
\eea

Efficiency of fast-roll inflation rapidly decreases 
once one considers the regime with $H \ll m$. An interesting 
example is provided by gauged N=8 supergravity, where, 
as we will see later, $|V''| = 2V$, i.e. $m^2 = 6H^2$. 
In terms of our  potential $V(\phi) = V_0 - {m^2\phi^2\over 2}$ 
this implies that the point $\phi_1$, which corresponds to 
$V(\phi_1) = V_0/2$, is given by $\phi_1 = 1/\sqrt 2 $. 
In dimensional units this is equivalent to having 
$\phi_1 \sim 1.5\times 10^{18}$ GeV.   In this model one
 has $F(m^2/H^2) = F(6) = 1.37$, and $1/F = 0.73$.
Then, for $m = \sqrt 6 \,H \sim 2\times 10^{-60}M_p$ 
one finds
\bea
e^{H\Delta t} \sim \left(5\  10^{60}\right)^{0.73} 
\sim {10^{44}} \sim e^{100}  \ .
\eea
Thus, fast-roll inflation in N=8 gauged supergravity 
may be responsible for up to 100 e-folds of exponential 
expansion of the universe with the Hubble constant similar 
to its present value $H \sim  10^{-60}M_p$.

On the other hand, for 
$m = \sqrt 6 H \sim   10^2$ GeV one has
\bea
e^{H\Delta t} \sim  {10^{13}} \sim e^{28}  \ .
\eea 

To give a different example, one may consider a 
simplest model of spontaneous symmetry breaking with 
the  potential
\bea
V(\phi) = {\lambda\over 4} (\phi^2-v^2)^2 =
 -{1\over 2} m^2\phi^2 + {m^2 \over 4 v^2} \phi^4 
+{1\over 4} m^2 v^2       \ ,
\eea 
where $m^2 = \lambda v^2$, and $\phi = v$ corresponds 
to the minimum of $V(\phi)$ with symmetry breaking. 
The Hubble constant at $\phi = 0$ in this model is given by 
$H^2 = {m^2 v^2\over 12}$, so that $F(m^2/H^2) = F(12/v^2)$. 
Fast-roll inflation in this model occurs for $\phi \lesssim v/2$.  
Assuming, e.g., $v = 1$ and $m \sim 100$ GeV,
 one finds $F^{-1}(12) = 0.44$, and 
\bea
e^{H\Delta t} \sim \left(10^{17}\right)^{1/F}  \sim 10^{7}\ ,
\eea 
whereas for $v = 0.7$, as in Polonyi model, one has 
\bea
e^{H\Delta t} \sim   10^{5}\ .
\eea

Thus, if in a certain class of theories one has
 $|V''| \sim V$, one should not immediately discard
 such theories as candidates for the description of 
an accelerated (inflationary) stage of the evolution 
of the universe. Such theories can describe a prolonged 
stage of fast roll inflation if $m  \ll 1$ in Planck 
units, see Eqs. (\ref{smallm1}), (\ref{smallm}). 
For $|V''| \sim V$, the requirement $m   \ll 1$  
 is equivalent to the requirement that the extremum 
of the effective potential corresponds to the energy 
density much smaller than the Planck density, $V \ll 1$.

A more detailed discussion of the fast-roll inflation will be contained in a 
separate publication \cite{fastroll}.

\section{\label{N=8}N=8 Gauged Supergravities}

\subsection{De Wit-Nicolai Potential}

The ungauged N=8 supergravity of Cremmer and Julia 
\cite{Cremmer:1979up} has a local $SU(8)$ symmetry and 
a rigid $SL(8, \mathbb{R})$ symmetry, equations of motion 
have a larger, non-compact $E_{7(7)}$ symmetry. The 70 real 
scalars of N=8 supergravity parametrize the coset space 
$E_7/SU(8)$ and can be described by an element 
${\cal V}(x)$ of the fundamental 56-dimensional 
representation
of $E_{7(7)}$:
\bea
{\cal V}(x)=
\left(
\begin{array}{cc}
u_{ij}^{\;\;IJ}(x) & v_{ijKL}(x)  \\
v^{klIJ}(x) & u^{kl}_{\;\;KL}(x)
\end{array} 
\right)\ .
\eea
Out of 133 fields 63 may be gauged using an $SU(8)$ symmetry.
De Wit and Nicolai \cite{deWit:1982eq} gauged the $SO(8)$ 
subgroup of $SL(8, \bf{R})$ symmetry of the ungauged 
supergravity. The $SO(8)$ gauge coupling constant $g$ 
is a new parameter which the gauged supergravity has, 
in addition to the gravitational constant. The local 
N=8 supersymmetry of gauged supergravity requires a 
non-trivial effective potential for the scalars. It 
is proportional to the square of the gauge coupling. 
The scalar and gravity part of de Wit-Nicolai action is   
\bea
\int d^4 x \sqrt{-g} \left( \frac{1}{2} R 
- \frac{1}{96} \left| A_{\mu}^{\;\;ijkl} \right|^2 - 
V \right),
\label{action}
\eea
where
the building blocks for the scalar kinetic terms are:
$
A_{\mu}^{\;\; ijkl} = -2 \sqrt{2}
 \left( u^{ij}_{\;\;IJ} 
\partial_{\mu} v^{klIJ} -
v^{ijIJ} \partial_{\mu} u^{kl}_{\;\;IJ} \right)
$.
De Wit-Nicolai nontrivial effective 
potential can be written as the difference
of two positive definite terms:
\bea
V= -g^2 \left( \frac{3}{4} \left| A_1^{\;ij} 
\right|^2-\frac{1}{24} \left|
A^{\;\;i}_{2\;\;jkl}\right|^2 \right), 
\label{V}
\eea
\bea
A_1^{\;\;ij} & =& 
-\frac{4}{21} T_{m}^{\;\;ijm}, \;\;\;
 A_{2l}^{\;\;\;ijk}=-\frac{4}{3}
T_{l}^{\;[ijk]},
\label{a1a2}
\eea
are some particular combinations of T-tensors:
\bea
T_l^{\;kij} & = & 
\left(u^{ij}_{\;\;IJ} +v^{ijIJ} \right)
 \left( u_{lm}^{\;\;\;JK} 
u^{km}_{\;\;\;KI}-v_{lmJK} v^{kmKL} \right).
\label{ttensor}
\eea

The 56-bein ${\cal V}(x)$ can be brought into
 the following form in the
$SU(8)$ unitary gauge by the 
 $SU(8)$ rotation
\bea
{\cal V}(x)=
\mbox{exp} \left(
\begin{array}{cc}
0 &  \phi_{ijkl}(x)  \\
 \phi^{ijkl}(x) & 0 
\end{array} \right),
\label{calV}
\eea  
where $\phi^{ijkl}$ is a complex self-dual tensor 
describing the 35 
scalars 
and 35 pseudo-scalars of $\phi^{ijkl}$)
  of $ N=8$ supergravity. The potential has an 
$adS_4$ critical point where all scalars and   
pseudo-scalars
vanish.

\subsection{Non-compact gaugings}
Compact gaugings of N=8 supergravity do not give 
de Sitter solutions, however the non-compact and 
non-semi-simple  gaugings with $CSO(p,q,r)$ groups, 
suggested and developed by Hull \cite{Hull:1984yy}, do have  de Sitter 
and Minkowski solutions. These are unitary 4d  theories 
with positive definite kinetic terms. One starts
 with the subalgebra of the $SL(8, \mathbb{R})$ 
algebra with the metric parametrized by two 
parameters $\xi$ and $\zeta$
\bea
\eta_{AB}
=\left( \begin{matrix} {\bf 1}_{p\times p} 
& \quad & \quad  \cr
\quad & \xi {\bf 1}_{q\times q} & \quad  \cr
\quad &  \quad & \xi 
\zeta{\bf 1}_{r\times r}   \end{matrix} \right)\ ,
  \qquad p+q+r=8.
\eea
It was shown in \cite{Cordaro:1998tx} that N=8 gauged supergravity 
can depend only on one 
continuous parameter that corresponds to a coupling constant 
and  parameters $\xi$ and $\zeta$ can have only descrete 
values 1,-1 and 0. The gauging for  the case 
$\xi=1,\, \zeta = -1$  corresponds to $SO(p+q,r)$ group, 
 $\xi= -1,\, \zeta = 1$ is $SO(p, q+r)$ gauging and so on. 
The $CSO(p,q,r)$ group is a group contraction of the
 $SO(p+r,q)$ group preserving a symmetric metric with 
$p$ positive eigenvalues, $q$ negative eigenvalues 
(for negative $\xi$), and $r$ zero eigenvalues. 

Recently the potentials that explicitly depend on two scalar fields $s$ 
and $t$  (out of 70 ) were proposed in \cite{Ahn:2001by}:
\bea
V_{\xi, \zeta}^{ p,q,r}( s, t )=g^2
\left( 4(\partial_{s} 
W_{\xi, \zeta}^{ p,q,r}( s, t))^2+
4(\partial_{t} 
W_{\xi, \zeta}^{ p,q,r}( s, t))^2-
6(W_{\xi, \zeta}^{ p,q,r}( s, t))^2
\right)
\label{pot2s}
\eea 
where the superpotential
 $W_{\xi, \zeta}^{ p,q,r}( s, t)$ is:
\bea
W_{\xi, \zeta}^{ p,q,r}( s, t)=
\frac{1}{8}
\left(pe^{\sqrt{\frac{4q}{p(p+q)}}s
-\sqrt{\frac{r}{2(p+q)}}t}+
q \xi e^{-{\sqrt{ \frac{4p}{q(p+q)}}s}
-\sqrt{\frac{r}{2(p+q)}}t}
+r\xi\zeta e^{\sqrt{\frac{p+q}{2r}}t}
\right)
\label{superpot2s}
\eea
and the kinetic terms for scalar fields have a canonical
 form. Gravitational and scalar part of the 
supergravity action for each model with $p,q,r$ 
and $\xi, \zeta$ is given by
\bea
S_{\xi, \zeta}^{ p,q,r}=\int d^4x\sqrt{-g}
\left( \frac{1}{2}R 
-\frac{1}{2}\partial^\mu  s
\partial_\mu  s 
-\frac{1}{2}\partial^\mu  t
\partial_\mu  t 
- V_{\xi, \eta}^{ p,q,r}( s, t )
\right).
\eea
Thus we have a family of models characterized by 
3 discrete parameters $p+q+r=8$ and by two parameters
 $\xi, \zeta$. At $r=0$, $p+q=8$ the meaning of  $\xi$
 can be inferred from the higher-dimensional 
interpretation of these models. It has been shown by 
Hull and Warner \cite{Hull:1988jw} that they can 
be obtained from 11d 
 supergravity (M-theory).
The general case of a compactification on 
a hyperboloid gives SO(p,q) gauging 
 and a compactification 
on a sphere leads to SO(8) gauging of 
de Wit and Nicolai \cite{deWit:1982eq}. 
The corresponding hypersurface constraining the
 internal 7-manifold is
\bea
r^2=\eta_{AB}z^Az^B
=\sum^p_{A=1}(z^A)^2 
+ \xi\sum^8_{A=p+1}(z^A)^2 .
\label{hypersurf}
\eea
For $\xi<0$ the expression (\ref{hypersurf})
represents a family of  hyperboloids and
for  $\xi>0$ it is a family of ellipsoids. 

 We have investigated the nature of the critical points
 corresponding to  de Sitter vacua for  potentials (\ref{pot2s}) 
from \cite{Ahn:2001by}. We have found that they are always 
saddle points, the eigenvalues of the matrix of the
 second derivatives have 
one positive and  one negative value. We have presented 
the results of the calculations in the Table 1 for $g^2=1$. 
\newpage

\bea
\begin{array}{|c|c|c|c|c|c|c|c|c|c|c|}
\hline 
  p & q & r &\xi &\zeta & s_{cr} & t_{cr}  
 & W(s_{cr},t_{cr}) 
& V(s_{cr},t_{cr})           & (m_1)^2     & (m_2)^2          \nonu \\
\hline
  1 & 2 & 5 &   &   &       & -0.7522  
&  \frac{1}{2} \times 3^{-3/8} &   &  10.53  & -5.264       \nonu \\
  1 & 4 & 3 &   &   &       &  0.7522  
& -\frac{1}{2} \times 3^{-3/8} &   &  -5.264 &  3.51        \nonu \\
  2 & 1 & 5 & 1 & -1& 0     & -0.7522  
&  \frac{1}{2} \times 3^{-3/8} &  
 2 \times 3^{1/4}  &  10.53  &  -5.264      \nonu  \\
  2 & 3 & 3 &   &   &       &  0.7522  &
  -\frac{1}{2} \times 3^{-3/8} &  & 
 -5.264 &   3.51       \nonu \\
  3 & 2 & 3 &   &   &       &  0.7522  & 
-\frac{1}{2} \times 3^{-3/8}  &  &  -5.264 & 
  3.51       \nonu \\
  4 & 1 & 3 &   &   &       &  0.7522  &  
-\frac{1}{2} \times 3^{-3/8} &  &  -5.264
 &   3.51       \nonu \\
\hline
  1 & 3 & 4 &   &   &       &          &   &  
 &  -4.    &  4.          \nonu \\
  2 & 2 & 4 & 1 & -1& 0     &  0       & 0                            & 2 &  -4.    &  4.          \nonu \\
  3 & 1 & 4 &   &   &       &          &  &                               &  -4.    &  4.          \nonu \\
\hline
  3 & 1 & 4 &   &   &0.4757 & -0.5826  & \frac{1}{2} \times 3^{-3/8}  &   &  -5.264 &  3.51        \nonu \\
  3 & 2 & 3 &   &   &0.6017 & -0.4513  & \frac{1}{2} \times 3^{-3/8}  &   &  -5.264 &  3.51        \nonu \\
  3 & 3 & 2 & -1& 1 &0.6728 & -0.3364  &\frac{1}{2} \times 3^{-3/8}   & 
                                                         2 \times 3^{1/4} &  -5.264 &  3.51        \nonu \\
  3 & 4 & 1 &   &   &0.7192 & -0.2202  & \frac{1}{2} \times 3^{-3/8}  &   &  -5.264 &  3.51        \nonu \\
  5 & 1 & 2 &   &   &-0.5014&  0.5606  & -\frac{1}{2} \times 3^{-3/8} &   &  10.53  & -5.264       \nonu \\
  5 & 2 & 1 &   &   &-0.6565&  0.367   & -\frac{1}{2} \times 3^{-3/8} &   &  10.53  & -5.264       \nonu \\
\hline
  4 & 1 & 3 &   &   &       &          &                              &   &   4.    &   -4.        \nonu \\
  4 & 2 & 2 & -1& 1 &  0    & 0        & 0                            & 2 &  -4.    &    4.        \nonu \\
  4 & 3 & 1 &   &   &       &          &                              &   &   4.    &   -4.        \nonu \\
\hline
  1 & 3 & 4 &   &   &-0.4757& -0.5826  & -\frac{1}{2} \times 3^{-3/8} &   & -5.264  &  3.51        \nonu \\
  1 & 5 & 2 &   &   & 0.5014&  0.5606  &  \frac{1}{2} \times 3^{-3/8} &   &  10.53  & -5.264       \nonu \\
  2 & 3 & 3 &-1 &-1 &-0.6017& -0.4513  & -\frac{1}{2} \times 3^{-3/8} & 
                                                      2 \times 3^{1/4}    & -5.264  &  3.51        \nonu \\
  2 & 5 & 1 &   &   & 0.6565& 0.367    & \frac{1}{2} \times 3^{-3/8}  &   &  10.53  & -5.264       \nonu \\
  3 & 3 & 2 &   &   &-0.6728& -0.3364  & -\frac{1}{2} \times 3^{-3/8} &   & -5.264  &  3.51        \nonu \\
  4 & 3 & 1 &   &   &-0.7192& -0.2202  & -\frac{1}{2} \times 3^{-3/8} &   & -5.264  &  3.51        \nonu \\
\hline
  1 & 4 & 3 &   &   &   &  &  &                                           &  4.     & -4.          \nonu \\
  3 & 4 & 1 &-1 & -1&   0   &    0     & 0                            & 2 &  4.     & -4.          \nonu \\
  2 & 4 & 2 &   &   &       &  &  &                                       & -4.     &  4.          \nonu \\
\hline
\end{array}
\nonu
\eea
Table 1. De Sitter critical points for
 $p,q,r$,  $\xi$, $\zeta$ models. One of the 
mass square eigenvalues $(m_1)^2$ or $(m_2)^2$ 
is negative, all critical points are saddles.

\newpage
\begin{figure}[b]
\centering\leavevmode\epsfysize=10cm \epsfbox{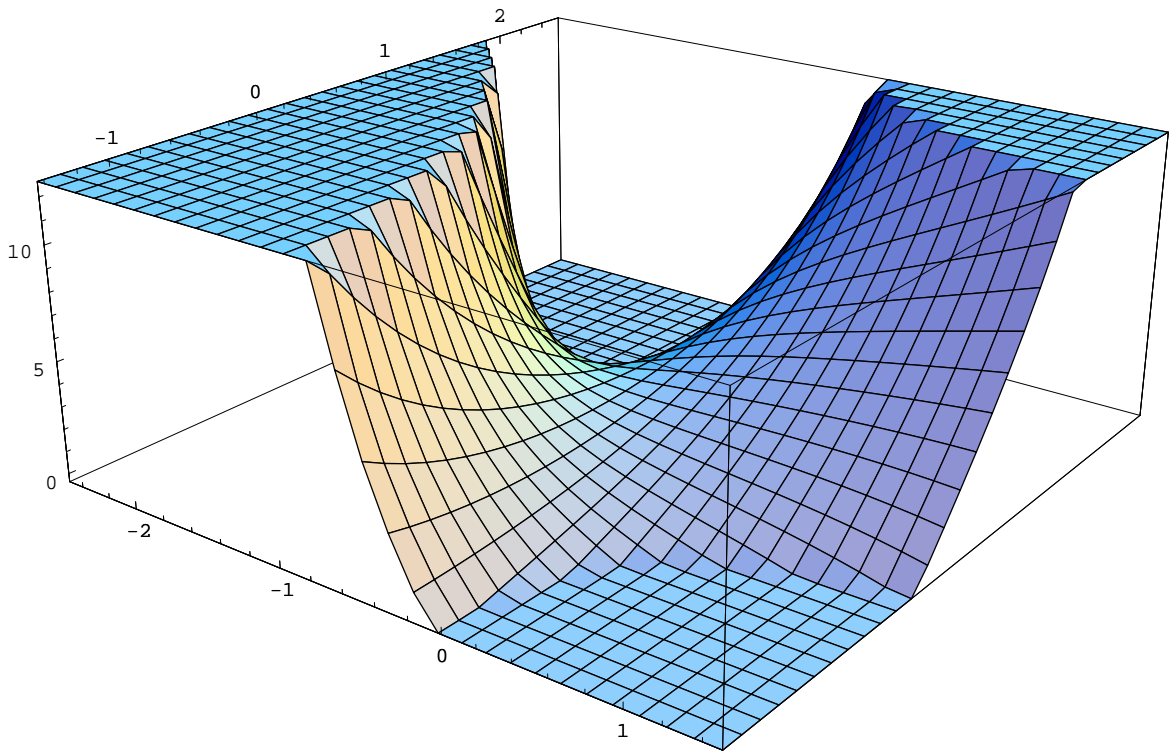}
\caption[fig1]{\label{saddle} An example of de
 Sitter saddle point in N=8 gauged supergravity.}
\end{figure}

For each model we give the critical values of the 
fields, a superpotential and a potential. The eigenvalues
 of the square of the mass matrix are $(m_1)^2$ and $(m_2)^2$. 
Their product is in all cases negative.
Now we may also compare the tachyonic mass near 
the critical point with the value of the potential 
there. We find that in all 27 cases a remarkable 
relation takes place:
$
-V''= |m^2|_{tach}= 2 V
$.
The absolute value of the square of the tachyonic 
mass is twice the value of the potential in units 
in which $M_p=1$. Using the curvature of the 
de Sitter space we can present these relations as
\bea
 |m^2|_{tach} = 2 V= {1\over2} |R|= 6 H^2 .
\eea
\noindent The positive square mass eigenvalues at
 the critical point also have a simple relation to 
the potential and/or to the curvature. In models 
where $W_{cr}=0$ one has
\bea
 m^2_{tach}= - {1\over 2} |R| \ ,
 \qquad  m^2_{pos}=  {1\over 2} |R|\ .
\eea
In models with  $W_{cr}\neq 0$ 
\bea
 m^2_{tach}= - {1\over 2} |R| ,  
\qquad  m^2_{pos}=   |R|\quad {\rm or}
 \quad m^2_{pos}= {1\over 3}  |R|\ .
\eea
Until now we described potentials  with an extremum 
corresponding to de Sitter  space. All of such extrema
 corresponded to saddle points of the type shown in 
Fig. \ref{saddle}. In all models of such type known so far,
 there is only one such saddle point, and the potential 
is unbounded from below, as we have checked. Whereas such
 potentials could play some role in the description of 
the present stage of exponential expansion of the universe,
 it is hard to see them playing any role in inflationary
 cosmology. The slow-roll condition is not satisfied,
 however, as we discussed in Sect. \ref{fast}, the fast-roll 
regime with $|V''|= 2V$ may be acceptable in certain cases. 
The situation may change once one considers potentials 
involving many other scalar fields and/or quantum 
corrections.

 Now we will also study the nature of the critical points
 for potentials depending on one field in the models with 
 $SO(p,q)$ gauging  with  $p+q=8$ and $r=0$ 
\cite{Hull:1984yy}. We will use the form of these models
 as given in \cite{Hull:1984yy},  with  a canonical kinetic 
term for the  the scalar field $\phi$ as in  \cite{Ahn:2001by}. 
The relevant part of the action of N=8 gauged supergravity is:
\bea
S_{p,q,\xi}=\int d^4x\sqrt{-g}
\left( \frac{1}{2}R 
-\frac{1}{2}\partial^\mu \phi
\partial_\mu \phi
-  V_{p,q,\xi}(\phi)
\right).
\eea
The potential  
can be  written with the help of  a 
superpotential $W_{p,q,\xi}$:
\bea
W_{p,q,\xi}
&=&
\frac{1}{8}\left(p
 e^{\sqrt{\frac{q}{2p}}\phi}
+q\xi e^{-\sqrt{\frac{p}{2q}}\phi}  
\right),
\label{suppot1}\\
 V_{p,q,\xi}
&=&g^2\left(
4(\partial_{\phi}
W_{p,q,\xi}(\phi))^2
-6W_{p,q,\xi}(\phi
)^2\right).
\label{pot1}
\eea
 In fact the de Sitter vacua here are the same as in 
 Table 1 where only one out of two scalars $s$ and $t$ 
are kept and the critical points correspond to a  maximum of 
the potential. There are three cases of de Sitter vacua.

\begin{itemize}
\item $p=q=4$ \bea
 V_{4,4,\xi} &=& -g^2\left(e^{\sqrt {2} \phi}
+4\xi +{\xi}^2 e^{-\sqrt {2} \phi}\right).
\label{pot44}
\eea
De Sitter critical point is at negative $\xi$ so that
$
 e^{-\sqrt {2} \phi_{cr}  }= \frac{1}{|\xi |}\ .
$
The value of the potential and of its second derivative at 
the critical points is:
\bea
 V_{4,4,\xi}|_{cr} = 
2 g^2|\xi |\ ,  \qquad 
 (V^{\prime \prime }_{4,4,\xi})_{cr} = -4g^2\,|\xi|\ .
\label{pot44cr}
\eea

\item $p=5$ , $q=3$ \bea
 V
 _{5,3,\xi} 
&=& -\frac{3g^2}{8}\left(5 e^{\sqrt{6\over 5} \phi}
+10 \xi e^{-\sqrt{2\over 15}\phi}
+{\xi}^2 e^{-\sqrt{10\over 3}
 \phi}\right) .
\label{pot53}
\eea
De Sitter critical point is at negative 
$\xi$ and 
$
 e^{-4 \sqrt {2\over 15} \phi_{cr}  }
= \frac{3}{|\xi |}\ .
$
The value of the potential and of its second 
derivative at the critical point is:
\bea
 V_{5,3,\xi}|_{cr} = 
2 g^2\cdot 3^{1\over 4}|\xi |^{3\over 4}\ ,
  \qquad 
 (V^{\prime \prime }_{5,3,\xi})_{cr} 
= - 4 g^2\cdot 3^{1\over 4}
|\xi |^{3\over 4} \ .
\label{pot53cr}
\eea

\item $p=3$ , $q=5$ \bea
 V_{3,5,\xi} 
= -\frac{3g^2}{8}\left(
e^{\sqrt{10\over 3}\phi}+10 
\xi e^{\sqrt{2\over 15} \phi}
+ 
5{\xi}^2 e^{-\sqrt{6\over 5}\phi} 
\right).
\eea
De Sitter critical point is at negative $\xi$ and 
$
 e^{-4 \sqrt {2\over 15} \phi_{cr}  }
= \frac{3}{|\xi |}\ .
$
The value of the potential and of its second 
derivative at the critical points is:
\bea
 V_{3,5,\xi}|_{cr} = 
2 g^2\cdot 3^{1\over 4}|\xi |^{5\over 4}\ , 
 \qquad 
 (V^{\prime \prime }_{3,5,\xi})_{cr} 
= - 4 g^2\cdot 3^{1\over 4}
|\xi |^{5\over 4} \ .
\label{pot35cr}
\eea

\end{itemize}

Thus  we find that in all 3 cases above the critical 
point is a maximum of the scalar potential and the 
tachyonic mass squared has  twice the value of
 the potential:
\bea
- V^{''}= |m^2|_{tach} 
= 2 V = {1\over 2} |R| \ .
\eea


\section{\label{N=4}de Sitter vacua of N=4 and N=2 
gauged supergravities }

The first de Sitter solutions of  gauged supergravity 
have been discovered by Gates and Zwiebach 
\cite{Gates:1983ct}, \cite{Zwiebach:1984qq} in the framework 
of $SU(2)\times SU(2)$ gauged version of the $SO(4)$ 
 N=4 theory.  It seems that SO(4) gauged supergravities 
have to have two independent gauge couplings $g_1$ 
and $g_2$ corresponding to each $SU(2)$. 
However, it was found by Zwiebach \cite{Zwiebach:1984qq} 
that it is not really the case 
due to the presence of scalar fields in front of 
the kinetic terms of the vector fields. These scalar fields
 acquire vacuum expectation values and this makes it
necessary to rescale vector fields and gauged couplings
so that the model has only one effective coupling constant.
For the case corresponding to a positive cosmological constant
$g_1$ and $g_2$ have to satisfy a relation
 $g_{1eff}= - g_{2eff}= \sqrt{g_1g_2}.$
It was shown that the value of cosmological constant 
 depends only on this one effective coupling constant.
 More general N=4 supergravities were studied in 
superspace in \cite{Gates:1983an}.

Later  de Roo and Wagemans  \cite{deRoo:1985jh} studied 
a more the general case of $SU(2)\times SU(2)$
gauging with separate phases 
$\alpha_{1,2}$ for each $SU(2)$ . 
For $\alpha=\alpha_{1}-\alpha_{2}= \frac{\pi}{2}$ 
the scalar potential proposed in \cite{deRoo:1985jh}
corresponds to the potentials from the papers of
Gates and Zwiebach \cite{Gates:1983ct},\cite{Zwiebach:1984qq}
and Hull \cite{Hull:1984yy}.
In all cases scalars parametrize the 
$SU(1,1)\over U(1)$ coset space. 

We will also take into account that some supergravity 
actions have the Einstein term as $\pm {1\over 2}R$,
 which corresponds to $\pm {M_p^2\over 2}R$ with $M_p=1$, 
whereas some other actions have $\pm {1\over 4}R$, 
which corresponds to $\pm {M_p^2\over 2}R$ with 
$M_p={1\over \sqrt 2}$. In N=8 case in the previous 
action we had $M_p=1$ case, in \cite{deRoo:1985jh} 
we have $M_p={1\over \sqrt 2}$ case. When we will 
compare the relation between tachyon mass and the 
potential, we will keep this in mind.
We will perform the study of de Sitter solutions, 
starting with the potential from de Roo and Wagemans 
paper \cite{deRoo:1985jh}:
\bea
V=-\frac{1}{2}\left( g_1^2|\Phi_1|^2
+g_2^2|\Phi_2|^2\right)
-ig_1g_2\left(\Phi_1^{\star}\Phi_2
-\Phi_2^{\star}\Phi_1\right),
\label{potRoW}
\eea
where scalar fields $\Phi_1$ 
and $\Phi_2$ are 
\bea
\Phi_1=e^{i\alpha_1}\phi^1
+e^{-i\alpha_1}\phi^2 \ ,
\nonumber \\
\Phi_1=e^{i\alpha_2}\phi^1
+e^{-i\alpha_2}\phi^2 \ ,
\nonumber
\eea
where  $\phi_1$ and $ \phi_2 $ 
are SU(1,1) doublet 
of scalar fields from N=4 Weyl multiplet 
\bea
 \phi^1 =(\phi_1)^{\star}\ ,
\qquad 
 \phi^2 =-(\phi_2)^{\star}\ ,
\label{phi}
\eea
with the constraint:
\bea
 \phi^a \phi_a = 1, \quad a=1,2 .\
\label{phiphi}
\eea
The solution of this constraint gives:
\bea
\phi_1
=\frac{e^{i\beta}}{\sqrt{1-|Z|^2}},
\quad \phi_2
=\frac{Ze^{i(\alpha_1+\alpha_2)}}
{\sqrt{1-|Z|^2}}\ ,
\eea
where 
$$e^{i\beta}
=\frac{e^{i\alpha}g_1^2
+e^{-i\alpha}g_2^2}
{|e^{i\alpha}g_1^2
+e^{-i\alpha}g_2^2|}$$
and $\alpha=\alpha_1-\alpha_2.$
The only remaining independent scalar field 
is $Z=X+iY$ and in terms of this field 
 $Z$ and parameter 
$\alpha=\alpha_1-\alpha_2$ the 
potential is:
\bea
V=-\frac{1}{2}\frac{1}{(1-|Z|^2)}
\left( (g_1^2+g_2^2)(1+|Z|^2)
-2|e^{i\alpha}g_1^2
+e^{-i\alpha}g_2^2|X\right)
-2g_1g_2\sin{\alpha} \ .
\eea
The critical point for this potential 
with the additional constraint $|Z|<1$, required 
for the positivity of the kinetic terms for scalars, is:
\bea
X_{cr} &=& Z_0  =   \frac{g_1^2+g_2^2-
2|g_1g_2 \sin(\alpha)|}
{|e^{i\alpha}g_1^2
+e^{-i\alpha}g_2^2|} \nonumber \\
&=&\frac{g_1^2+g_2^2-
2|g_1g_2 \sin{\alpha}|}
{\sqrt{(g_1^2+g_2^2-
2g_1g_2 \sin{\alpha})(g_1^2+g_2^2+
2g_1g_2 \sin{\alpha})}}\ ,\nonumber \\
Y_{cr}&=&0 .\nonumber
\eea
The potential at this point is 
$
V|_{cr}
=-|g_1g_2 \sin{\alpha}|-2g_1g_2 \sin{\alpha}\ .
$
For  $2g_1g_2 \sin{\alpha}<0$ the potential 
is positive
\bea
V_{cr}
=|g_1g_2 \sin{\alpha}|.
\eea
To find a second derivative of the potential we have 
to find the scalars which have  canonical kinetic 
terms at the critical point.
Using (\ref{phi}) and (\ref{phiphi})
 we will get kinetic terms at the critical point:
\bea
\frac{1}{2}\partial\phi^a \partial\phi_a
=  - \frac{1}{2} \frac{1}{(1 - Z_0^2)^2}
\partial X\partial X 
 - \frac{1}{2} \frac{1}{(1 - Z_0^2)}
\partial Y\partial Y.
 \eea
Thus $x= \frac{1}{(1 - Z_0^2)}X$ and 
$y= \frac{1}{(1 - Z_0^2)^{1/2}}Y$ are the fields
 over which we have to differentiate the potential.
We find:
\bea
m^2_{xx}
&=&-4|g_1g_2 \sin{\alpha}|\ , \\
m^2_{yy}&=&-(g_1^2+g_2^2+
2|g_1g_2 \sin{\alpha}|).
\eea
Thus we see  (with account of  normalization  
$M^2_p= {1\over 2}$ in \cite{deRoo:1985jh}) that 
 the tachyonic mass in the $x$ direction compares 
with the potential as in all previous cases:
\bea
-M^2_p V_{cr}^{''}= 2 V_{cr}\ , 
\qquad m^2_{tach}= -{1\over 2}|R|\ .
\eea
However, the potential in $y$ direction has  a tachyonic 
mass which looks quite independent of the value of the 
potential at the critical point.
This looks puzzling in view of the large amount of 
examples considered before. However, in a 
particular case of $\alpha = {\pi\over 2}$ this 
puzzle is resolved as follows: 
\bea
V_{cr}
=|g_1g_2|\ , \qquad m^2_{xx}
= -4|g_1g_2| \ , \qquad 
m^2_{yy}=-(g_1+g_2)^2 \ .
\eea
This is still puzzling, however here we have to remember
 that without looking at kinetic terms for vector fields,
 one can not make a definite judgement about the relation 
between $g_1$ and $g_2$. But this analysis was performed 
by Zwiebach in \cite{Zwiebach:1984qq} and he concluded that 
effectively one has to consider only the case of $g_1=-g_2$ 
for the de Sitter solution. This case gives us $m^2_{yy}=0$. 
In fact the potential depends only on the combination
 $Z^2=X^2+Y^2$ and there is only one  tachyon excitation 
and a flat direction. Thus suggests that if we would 
perform the analysis of the kinetic terms for vector
 fields for the theory with $\alpha \neq {\pi\over 2}$ 
we would find again that for canonical kinetic terms
 we do not have an extra tachyon field.

Another form of potential of N=4 theory was given in 
 \cite{Hull:1984yy}
and has been recently discussed in 
\cite{Hull:2001ii}:
\bea
V = - \left(4g_1g_2+
(g_1^2+g_2^2)\cosh{(a|\varphi |)}
+(g_1^2-g_2^2)
\frac{Re \varphi}{|\varphi|}
\sinh{(a|\varphi|)}\right)
\eea
with $W= {\varphi \over |\varphi |} 
\tanh{(\frac{a |\varphi |}{2})}.$
For $g_1=-g_2$ we get the potential 
discussed in \cite{Hull:2001ii} 
$$V=-\frac{1}{2} g^2
(\cosh{(a |\varphi |)} -2) $$
with $g^2=-4g_1g_2.$ The critical 
point is $\varphi =0$
and $V|_{cr}=\frac{1}{2} g^2.$
The presence of the parameter $a$ 
will not affect the properties of the system
even though it seems that in this case 
$V^{\prime\prime}|_{cr}= -\frac{a^2}{2} g^2$
and for very small $a$ it is possible to get 
 $V|_{cr} \gg |V^{\prime\prime}|_{cr}.$
The properties of the potential are related to
 canonical kinetic terms and the rescaling of the scalar
 field  $\varphi $  in the potential will also lead to 
the rescaling of the kinetic terms therefore we
 conclude that there is no
free adjustable parameter which can be used for a
 slow-roll condition.

The $ SU(2)\times SU(2)$ gauging
of N=4 supergravity can be easily reduced
to N=2 gauged supergravity with one vector 
multiplet gauging \cite{deWit:1984pk}. In this 
case $g_1=-g_2$ and the potential has a form:
\bea
V=2g_1^2-4g_1^2\frac{|Z|^2}{(1-|Z|^2)} \ .
\eea
The critical point  corresponds to  $Z=0$
and $V|_{cr}=2g_1^2.$  Again in this case we find that
 the tachyon mass is related to the potential as 
$- M_p^2 V^{''}_{cr}= 2 V_{cr}$.

To summarize, in all cases of N=4 and N=2 gauged supergravities which we looked at, 
we find tachyons with relation to the curvature of 
the de Sitter space of the form
$m^2_{tach}= -{1\over 2}|R|$. The situation for gauged 
N=2 supergravities may be more complicated in general and required 
additional consideration.


\section{\label{flat} Minkowski vacua and
 a possibility of inflation along flat directions}

There is another class of potentials which should 
not be overlooked in our search for de Sitter 
solutions. For certain values of parameters the
 potentials have flat directions corresponding 
to Minkowski space with $V(\phi) = 0$.
 Existence of Minkowski or near-Minkowski
 ground state is a pre-requisite of a
 successful inflationary cosmology, so even 
though at the classical level such potentials 
do not have any de Sitter solutions, their
 existence is rather intriguing.

Several examples of such models have been 
presented in \cite{Ahn:2001by}. The corresponding 
results can be summarized by the following table.

\bea
\begin{array}{|c|c|c|c|c|c|c|c|c|c|c|}
\hline 
  p & q & r &\xi &\zeta & s_{cr} & t_{cr}  
  &      W(s_{cr},t_{cr})   
   &V(s_{cr},t_{cr}) & m^2_1  
   & m^2_2          \nonu \\
\hline
  1 & 1 & 6 & 1 & 0 & 0   
  &{\rm any} &{1\over 4}
\exp{\left[-\sqrt{3\over2}t_{cr}\right]}&0 & 
 2\,e^{-\sqrt 6\,t_{cr}}   &   0  
  \nonu \\
\hline
  2 & 1 & 5 &   &   &       &     
     &                          
    &   &   0     &   0    
      \nonu \\
  2 & 2 & 4 &   &   &       &          &         
                     &   &   0     &   0   
       \nonu \\
  2 & 3 & 3 & 0 &1,0,-1&
{\rm any}&{\rm any}&{1\over4}
\exp{\left[{2s_{cr}
-\sqrt{5}t_{cr}}\over\sqrt{6}\right]}
&0&0&0\nonu \\
  2 & 4 & 2 &   &   &       &       
   &                            
  &   &   0     &   0       
   \nonu \\
  2 & 5 & 1 &   &   &       &   
       &                        
      &   &   0     &   0    
      \nonu \\
\hline
\end{array}
\nonu
\eea
The class of theories in this table has a possibility
 to be related to M-theory avoiding the non-compactness 
problem. In all cases here $\xi\geq 0$, thus there is 
no negative components of the metric on the hypersurface
 (\ref{hypersurf}). Instead, there are some
 non-compact $U(1)$ directions, for example,
 in $p= 1$, $q = 1$, $r = 1$, $\xi = 1$, 
$\zeta = 0$ case the internal space is 
$\mathbf{S}^2\times \mathbf{R}^6$ and
 the symmetry is
$SO(2)\times U(1)^{15}$.
However the flat directions $\mathbf{R}^6$
 could be compactified to $\mathbf{T}^6$ as 
explained in \cite{Hull:2001ii}.

 In the case $p= 1$, $q = 1$, $r = 1$, $\xi = 1$,
 $\zeta = 0$ the potential of the scalar 
fields $s$ and $t$ looks as follows:
\begin{equation}
V(s,t) = \frac{1}{8}\ {e^{-{\sqrt{2}}\,
  (2s+{\sqrt{3}}t )}}\
 \left(1-{e^{2{\sqrt{2}}\,
 s}}\right)^2 = \frac{1}{2}\ 
 e^{- \sqrt{6} t } 
\ \sinh^2 \,(\sqrt {2 }\,s)\ .
\end{equation}

\begin{figure}[b]
\centering\leavevmode\epsfysize=9cm 
\epsfbox{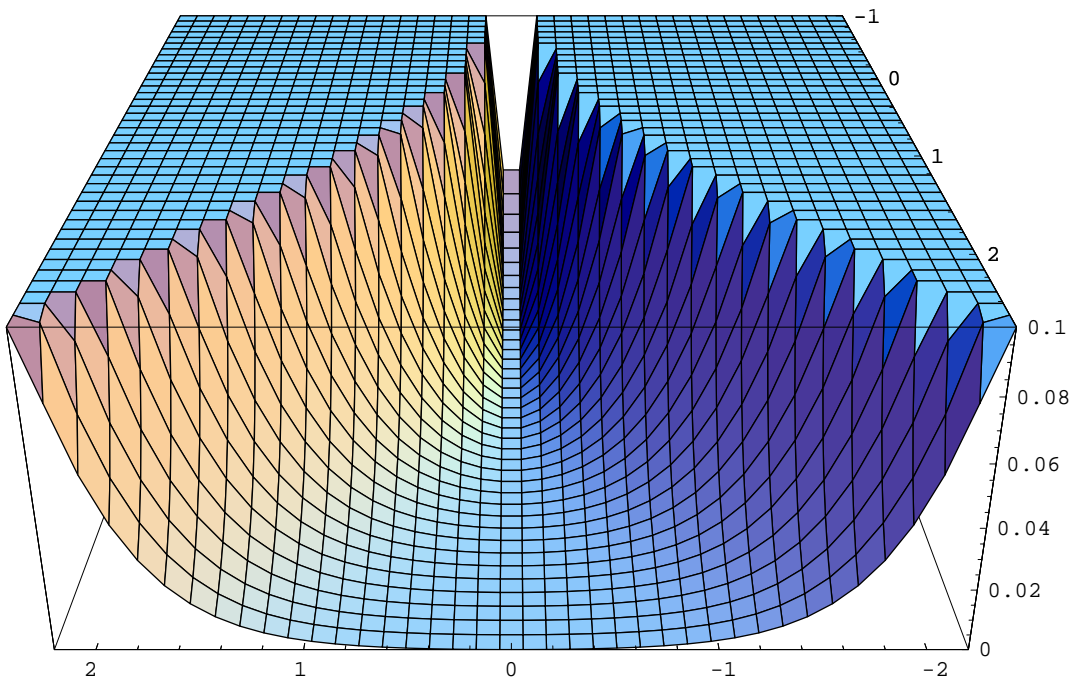}
\caption[fig1]{\label{valley} A potential with a valley 
with $V = 0$ corresponding to  Minkowski space.}
\end{figure}

This potential blows up at large  negative $t$  and at
 large $|s|$, it is even with respect to $s$, and it  
vanishes for all $t$ along the valley $s = 0$, 
see Fig. \ref{valley}. Its curvature is positive
 in all directions, no tachyons are present.

This potential is not of inflationary type, at least at 
the classical level. Indeed, there is no slow-roll regime 
of inflationary type  in the theories with exponential
 potentials of type $e^{c\phi}$ with $c \geq 1$; such 
potentials are too steep. Meanwhile,   the potential
 along the valley vanishes, so  inflation does not 
appear for the fields moving along  this valley as well. 
One could hope that the regime of acceleration may emerge 
when the fields oscillate in the valley and slowly drift 
towards positive $t$. However, the energy density of the
 oscillating field rapidly drops down. We have checked 
numerically, for various initial conditions for $s$ and $t$, 
that the cosmological evolution in the theory with this 
potential is not inflationary.

A similar situation appears in the model with $p = 2$, 
described in the lower part of the table. In this model 
the potential  vanishes (i.e. the cosmological constant 
is equal to zero) for all   $s$ and $t$. Once again, in 
this case one does not have an inflationary regime.

However, it might happen that the flat directions  can be
 lifted  due to quantum effects. Note that in these models 
supersymmetry is broken along the  flat directions, 
so it does not protect the effective potential against 
radiative corrections.

In this respect it is instructive to remember the recent 
example related to P-term inflation in   N=2  
 \cite{Kallosh:2001tm}, as well as  very similar examples
 of D-term inflation \cite{Dvali} and F-term inflation
 in N=1 \cite{Lyth}. The effective potential in P-term
 inflation at the classical level has several Minkowski
 flat directions with unbroken supersymmetry. 
Perturbative effects cannot lift up these flat
 directions and give rise to inflation.
 However, once one takes into account possible FI terms,
 these flat directions are lifted up to a state
 with $V >0$. They still remain flat, but they 
correspond to de Sitter vacuum with broken supersymmetry.
 Then the  radiative corrections, which appear because 
of the supersymmetry breaking,  make the effective
 potential $V$ slightly tilted. This leads to a 
realization of the hybrid inflation scenario in N=2.

We do not know whether anything like that will happen 
in N=8, but this is a very interesting possibility that 
deserves  separate investigation. 

One should note also that in  this paper we concentrated on the
 investigation of potentials with extrema at finite values 
of the scalar fields. However, one may also look for 
possible classical potentials that may have flat
 directions approaching de Sitter state with $V >0$ 
or Minkowski state with $V = 0$ {\it only asymptotically}.
 From the point of view of cosmology such potentials
 are very interesting: They can describe inflation 
at large $V(\phi)$ and quintessence at small $V(\phi)$.


\section{Conclusions}

In this paper we investigated the possibility to have de
 Sitter-like solutions describing inflation/accelerated 
expansion of the universe in some
 N=8,4, 2  gauged supergravities. In each model 
that we have studied we
have found that de Sitter state corresponds to a 
single unstable extremum of the scalar potential
 $V$. The (negative) curvature of the
potential in the direction of the fastest descent
 in all of these models
obeys the simple rule: $|V''| = 2V$ in  units $M_p = 1$.  
(Note that throughout the paper we are using the 
following definition of the Planck
mass: $M_p^2 = 8\pi/G$, where $G$ is the gravitational 
coupling
constant.)  This relation can be represented in the 
following way:
$|m^2| = |R|/2$, where $m^2$ is the tachyonic 
mass corresponding to the excitation in the 
direction of the fastest descent, and $R$ is 
the curvature scalar, $|R| = 12 H^2$.
It would be very interesting to find a simple 
geometric explanation of this relation. More general relations between tachyonic $|m^2|$ and a potential may be possible, particularly in $N=2$ theories \cite{Priv}. In this paper 
we concentrated on the derivation of this relation for 
a large class of models, and on investigation of its
consequences.

One of the consequences is that the slow-roll
 inflationary regime is
impossible near the extrema of the scalar potential 
in such models.
Note, however, that this conclusion may change when 
one takes into account  more scalar fields and more general gaugings. In N=2 gauged supergravities with vector and hypermultiplets one may expect some new results.
 We have studied only those models of N=8,4,2 supergravities for 
which an explicit expression for the potential
 was known. In these models the potential depends 
only on one or two scalar fields. Thus it may happen 
that some of our conclusions are not generic. 
It would be interesting, in particular, to look 
for potentials that have flat directions with 
 $|V''| \ll V$ asymptotically approaching  
Minkowski vacuum with $V =0$  or de Sitter
 vacuum with $V \lesssim 10^{-120}$, which 
would correspond to the present state of the universe.

If such a regime is possible, one may start looking 
for it already at the classical level. One may also
 try to find out whether the
flat directions with $V = 0$, which are present 
in some versions of  $N = 8$ gauged supergravity, 
can be lifted up by quantum effects, and can
 be used for implementation of inflation. An
 example of how this
possibility could be realized in N= 2 have been 
recently given in
\cite{Kallosh:2001tm}; see also examples 
in N= 1 given in \cite{Dvali,Lyth}.

On the other hand, if one is only interested 
in describing the present stage of quasi-exponential 
expansion of the universe in a state with
 $V \sim 10^{-120}$, the danger of the vacuum 
instability can be removed to a very distant 
future, hundreds of billions of years from the
 present epoch: The universe with
 $|V''| = 2V \sim   10^{-120}$ may experience 
more than 100 e-folds of fast-roll inflation, 
which is more than sufficient to explain the present
 stage of accelerated expansion. Of course, such 
models require rather extreme values of parameters,
 but it is still interesting that such a regime is
 possible despite the expected strong 
instability of de Sitter space with $|V''| = 2V$.

\subsection*{Acknowledgments}
The authors are grateful to G. Dall'Agata, G. Felder, P. Fr\'{e},  C. Hull, N. Kaloper, S. Kachru, L. Kofman, P. Townsend, A. Van Proeyen, and E. Witten  for useful discussions.  This work  was
supported by NSF grant PHY-9870115.  The work of M.S. was partly supported by  the Department of Energy under a contract DE-AC03-76SF00515. S.P. acknowledges the support from Stanford Graduate Fellowship Foundation.

\subsection*{Note Added} While we were preparing this paper for the submission we became aware of a related investigation by Paul Townsend, hep-th/0110072. He has found a model based on N=8 gauged supergravity with a sufficiently flat potential, which may lead to a marginally  inflationary regime $a(t) \sim t^3$. Such potentials can be useful for the description of the present acceleration of the universe, as proposed in hep-th/0110072. It would be interesting to use such potentials for a description of inflation in the early universe. However,  density perturbations produced in the universe with  $a(t) \sim t^3$ have a substantially non-flat spectrum, $n$ is well below 1, whereas observations suggest $|n-1| \lesssim 0.1$.  
One should note also that the model proposed in hep-th/0110072 is based on finding de Sitter solution in d=5 and making a subsequent reduction to d=4. In order to check whether this model is realistic it would be important to find out whether the corresponding 5d de Sitter solution may suffer from the same problem of instability as all 4d de Sitter solutions analysed in our paper. 

\

\

\end{document}